\documentclass[conference,10pt,letterpaper]{IEEEtran}

\usepackage{graphicx} 
\usepackage[hidelinks,bookmarks=false]{hyperref}
\usepackage{booktabs}
\usepackage{xcolor}
\usepackage{amsmath, amssymb}
\usepackage[linesnumbered,ruled,vlined]{algorithm2e}
\usepackage{algpseudocode}
\usepackage{subcaption}  
\usepackage{comment}

\usepackage{graphicx} 

\title{Adaptive AI Model Partitioning over 5G Networks}
\author{
\IEEEauthorblockN{Tam Thanh Nguyen, Tuan Van Ngo, Long Thanh Le, Yong Hao Pua,\\
Mao Van Ngo, Binbin Chen, and Tony Q.~S.~Quek} 
\IEEEauthorblockA{{\small \{{nguyen\_thanhtam}, {vantuan\_ngo}, {thanhlong\_le}, {vanmao\_ngo}, {yonghao\_pua}, binbin\_chen, tonyquek\}@sutd.edu.sg}}} 
\date{March 2025}

\begin{document}
\maketitle
\begin{abstract}

Mobile devices increasingly rely on deep neural networks (DNNs) for complex inference tasks, but running entire models locally drains the device battery quickly. Offloading computation entirely to cloud or edge servers reduces processing load at devices but poses privacy risks and can incur high network bandwidth consumption and long delays. Split computing (SC) mitigates these challenges by partitioning DNNs between user equipment (UE) and edge servers.
However, 5G wireless channels are time-varying and a fixed splitting scheme can lead to sub-optimal solutions. 
This paper addresses the limitations of fixed model partitioning in privacy-focused image processing and explores trade-offs in key performance metrics, including end-to-end (E2E) latency, energy consumption, and privacy, by developing an adaptive ML partitioning scheme based on real-time AI-powered throughput estimation. Evaluation in multiple scenarios demonstrates significant performance gains of our scheme.

\end{abstract}

\section{Introduction}



The proliferation of mobile devices and their growing reliance on deep neural networks (DNNs) for tasks such as image recognition~\cite{simonyan2014vgg}, natural language processing (NLP), and generative AI has introduced significant challenges in balancing computational efficiency, latency, energy consumption, and data privacy. While mobile hardware continues to improve, the execution of full-scale DNN inference on UE remains resource intensive and drains battery quickly. To mitigate this burden, computation is often offloaded to cloud or edge servers. Although this approach reduces processing demands on the device, it introduces new concerns regarding data privacy and dependence on network quality.

Split computing (SC)~\cite{neurosurgeon, boomerang} has emerged as a promising solution, offering a middle ground by partitioning DNN models between the UE and nearby edge servers. This technique enables partial processing on the device and delegates the remaining inference to the edge, thereby reducing local resource consumption while limiting the extent of data exposed beyond the device. Despite its advantages, most existing SC approaches rely on static partitioning strategies that fail to adapt to the dynamic nature of wireless environments. In 5G networks, rapid fluctuations in channel throughput and latency—exacerbated by interference from neighboring base stations and multiple competing UEs—introduce significant variability in network performance. Furthermore, UE mobility and environmental transitions, such as moving between indoor and outdoor areas, vehicular movement, and changes in line-of-sight conditions, result in unstable connectivity and variable signal quality. Compounding these issues are potential jamming attacks, including barrage jamming, which targets entire frequency bands, and smart jamming, which selectively disrupts specific data streams. Under such volatile conditions, fixed model partitioning leads to sub-optimal resource utilization, increased E2E latency, and a degraded quality of experience for the end user.

To address these limitations, this paper investigates adaptive ML model partitioning for privacy-sensitive image processing applications in 5G networks. We propose a novel framework that leverages real-time AI-powered spectrum sensing to dynamically adjust the DNN partitioning point based on current channel conditions. Using the AI-RAN platform with NVIDIA Aerial 5G stack, we predict the maximum achievable throughput for each UE and optimize ML model splitting accordingly. Our approach systematically evaluates the trade-offs among E2E latency, energy consumption, throughput, and privacy, demonstrating that adaptive partitioning significantly outperforms static schemes across key performance indicators. Our main contributions are:

\begin{itemize}
    \item An adaptive ML-partitioning scheme, considering privacy, E2E delay, and UE energy consumption jointly\footnote{Our prototype was demonstrated in AI-RAN Alliance booth at Mobile World Congress (MWC) Barcelona 2025, see \href{https://youtu.be/OXt4shUh0fs}{https://youtu.be/OXt4shUh0fs} }.
    \item A novel ML architecture to capture both the temporal relation of radio measurement metrics and In-phase and Quadrature (IQ) data as additional features, to predict maximum achievable throughput under diverse dynamic wireless channels.
    \item The proposed solution was evaluated on a 5G testbed based on NVIDIA Aerial stack. Compared to baselines, our scheme achieved an improvement of up to 65\% in E2E delay, while preserving privacy and balancing energy consumption of UE.
\end{itemize}

\section{Related Works}
\label{sec:relatedWork}

\subsection{ML model Splitting in Edge-Cloud Networks}
Neurosurgeon \cite{neurosurgeon} introduced an automated approach for splitting DNN model by dynamically determining the best partition point, based on estimation of latency and energy consumption of each DNN layer on both UE and server. 
Eshratifar \textit{et al.}~\cite{jointDnn} proposed JointDNN, 
considering multiple constraints, including battery limitation, cloud server congestion, and quality of service, to optimize mobile energy consumption or latency at run time.
Li \textit{et al.}~\cite{jalad} presented JALAD, 
which employs in-layer feature map compression and adjusts the splitting point based on network conditions and device capabilities to minimize execution latency while preserving model accuracy.
Banitalebi \textit{et al.}~\cite{auto-split} proposed Auto-Split, 
applying UE-cloud splitting and post-training quantization to the UE DNN. Auto-Split considers several constraints related to UE devices, networks, and cloud devices to determine the optimal splitting point.
Zeng \textit{et al.} \cite{boomerang} presented Boomerang, 
leveraging DNN right-sizing, an early-exit mechanism, along with DNN partitioning to optimize latency and accuracy within the available bandwidth constraints.
Although prior works have made significant contributions to model splitting in UE-cloud networks, they each focus on optimizing a single performance metric at a time, leaving room for a more holistic approach that jointly optimizes multiple metrics. Moreover, privacy concerns regarding offloading personal data to edge/cloud servers remain unaddressed. In Section \ref{sec:SystemArchitect}, we present our system architecture, which aims to bridge these gaps by providing a more comprehensive optimization strategy through privacy-aware model partitioning and multi-metric optimization.

\subsection{Throughput Estimation Techniques for Dynamic Networks}
\label{subsec:RelatedWorkThroughputEstimation}

Elsherbiny \textit{et al.}~\cite{Elsherbiny_Globecom2020_4G_throughput} studied 4G Long-Term Evolution (LTE) throughput prediction 
using \textit{client-side metrics} including Reference Signal Received Power (RSRP), Received Signal Strength Indicator (RSSI), Signal-to-Noise Ratio (SNR). They noted coarse granularity (1 second) as a limitation. In our work, an xApp receives data from Near-RT RIC every 0.1 seconds and feeds it into the ML-based throughput estimator.
Minovski \textit{et al.}~\cite{Minovski_Thrput_TMC_2023} presented a comprehensive study on throughput prediction using ML in LTE and non-standalone (NSA) 5G networks. Their work leverages lower-layer radio metrics (e.g., RSSI, RSRP, SINR) to develop non-intrusive ML models, achieving prediction accuracies of 93\% (R$^2$) for LTE and 84\% (R$^2$) for 5G. 
However, this study is limited to NSA 5G and does not consider more challenging environments such as under jamming or interference, which our work will address to fill the gap and to have a more robust estimation.
Besides the radio measurement metrics as prior works, we observed that these metrics could fail to estimate throughput under challenging environments, as discussed in Section~\ref{sec:ThroughputEstimation}. We realize that IQ data from Physical (PHY) layer and other radio measurement metrics help boost accuracy of throughput estimation for both normal and challenging wireless environments.
IQ data has been used for anomaly detection, Zhou \textit{et al.}~\cite{zhou2021radio} proposed AI-based radio anomaly detection by processing IQ-based spectrogram images.
However, the evaluation relies on simulated Orthogonal Frequency-Division Multiplexing (OFDM) data, which lacks the complexity of real 5G signals with multiple physical channels. 
For a challenging environment, Sciancalepore \textit{et al.} \cite{sciancalepore2023jamming} addressed the case of disrupting drone communication under jamming, where early detection of jamming is very important. Instead of observing consequent metrics (e.g., high 
BER, high SNR, or low Packet Delivery Rate--PDR) as post jamming, the authors used IQ samples from raw PHY layer, converted into black-white images, to detect jamming early in low-BLER, 
enabling early detection before link failure occurs.
Varotto \textit{et al.} \cite{varotto2024detecting} proposed an ML-based monitoring watchdog device under the 5G system to detect narrowband jammers, specifically targeting synchronization signal blocks (SSB). 
While achieving high detection accuracy with ML-based model with spectrograms from IQ data, the study only trained and tested with limited scenarios, and required deploying a monitoring device to detect jammers. In contrast, we introduce built-in spectrum sensing inside gNB to estimate throughput under various challenging wireless scenarios. 

\section{Adaptive ML Model Partitioning}
\label{sec:SystemArchitect}

\begin{figure}[!t]
    \centering\includegraphics[width=1\linewidth]{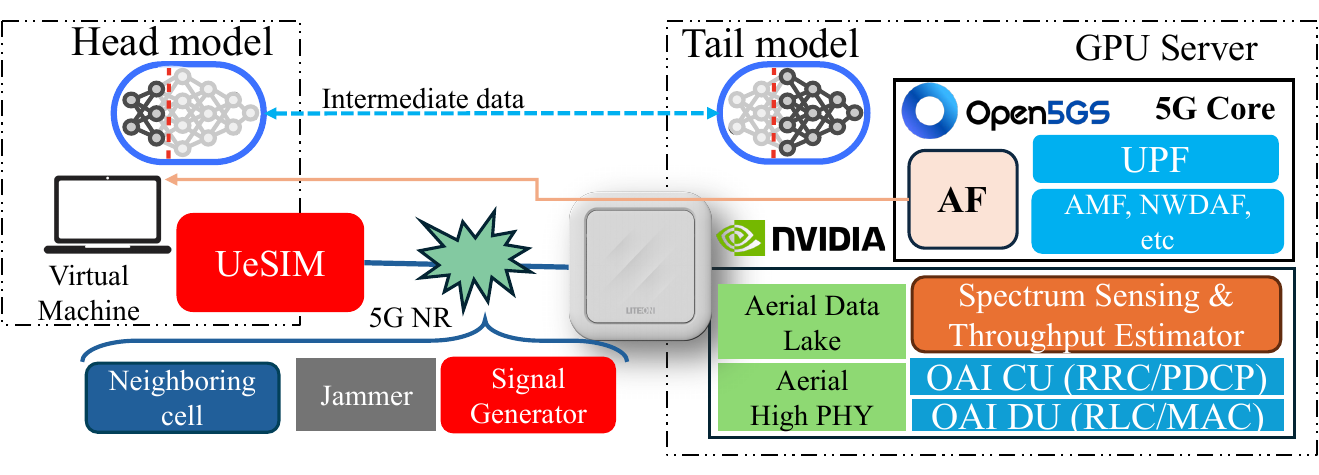}
    \vspace{-5mm}
    \caption{System Architecture of Adaptive AI Model Partitioning over dynamic 5G channels with NVIDIA Aerial 5G Stack.}
    \vspace{-3mm}
    \label{fig:SystemArchitecture}
\end{figure} 

Fig.~\ref{fig:SystemArchitecture} shows the system architecture of Adaptive AI model partitioning over dynamic 5G channels with NVIDIA Aerial 5G stack. 
The Deep Neural Network (DNN) is partitioned into two components at the layer $l$, where:
\begin{itemize}
    \item Head model (layer 1 to $l$) is executed on the UE.
    \item Tail model (layer $l+1$ to $L$) is offloaded to Edge server.
\end{itemize}

During inference, intermediate features extracted by the head model are transmitted over a 5G network, which operates in a dynamic 5G wireless environment. To optimize performance under these conditions, the model can be dynamically partitioned at an optimal splitting point $l^*$, determined based on recommendations from the \textit{Application Function (AF)}, which leverages real-time network insights gathered from gNB, to ensure efficient and adaptive operation.

We utilize the 5G gNB---NVIDIA Aerial CUDA-Accelerated RAN, a commercial-grade, software-defined, and cloud-native framework designed for the deployment of 5G and future 6G radio access networks (RAN). This framework enables the concurrent execution of both AI inference and RAN processing on a shared GPU server, thereby optimizing resource utilization.

Additionally, we integrate the Aerial Data Lake, a data capture platform that facilitates the collection of radio frequency (RF) data from the Aerial CUDA-Accelerated RAN. IQ samples from O-RUs, along with associated channel information, are transmitted to the host CPU and subsequently exported to the Aerial Data Lake database for further analysis.

Building on this robust infrastructure, we developed the RAN Channel Sensing and Estimator, a critical tool that provides real-time insights into 5G signal conditions, which are leveraged by the \textit{AF} to drive dynamic optimization.
We will explore its adaptive decision-making process in greater depth later.

To simulate various channel conditions in a 5G network, we introduce three types of interference sources.
(i) A Keysight signal generator is used to produce different jamming signals.
(ii) An additional 5G base station is deployed as a neighboring cell, generating inter-cell interference to the target base station.
(iii) A neighboring UE is also configured to cause uplink interference to the same base station.
Experiments are conducted in both over-the-air (OTA) and conducted modes. In OTA testing, the setup is enclosed within a Faraday cage to isolate it from external interference. In conducted mode, interference signals are directly injected into the RF path using splitters, combiners, and attenuators placed between the UE and the gNB.

\section{AI-based Throughput Estimator}
\label{sec:ThroughputEstimation}

To decide which splitting point between UE and server is the best, the maximum achievable throughput of UE to upload data via 5G network is an important parameter. 
We propose an AI-based throughput estimator that combines both numerical data of time series and image-based spectrogram.


\subsection{Key Performance Metrics (KPMs) from previous work}
Minovski \textit{et al.} \cite{Minovski_Thrput_TMC_2023} used seven relevant radio measurements to estimate the guaranteed throughput for 5G NR: (1) RSRP, (2) RSRQ, (3) SINR, (4) P\_a (Power Adjustment), (5) RI (Rank Indicator), (6) CQI (Channel Quality Indicator), (7) CRI (CSI-RS Resource Indicator). Unlike \cite{Minovski_Thrput_TMC_2023}, we provide seamless integration by collecting these metrics on the gNB side, which does not require the UE to have additional software.

\subsection{Our proposed additional Key Performance Metrics (KPMs)}


Besides above KPMs, we consider eight more KPMs: (1) PUSCH-SINR/PDSCH-SINR, (2) Transmission Power Control (TPC), (3) UL-MCS/DL-MCS, (4) UL-BLER/DL-BLER, (5,6,7,8) PUSCH-HARQ-(0,1,2,3)/PDSCH-HARQ-(0,1,2,3). 
Even including these additional numerical KPMs, we still observe that these metrics cannot well characterise the UE's UL throughput, especially under interference scenarios. To fill the gap, we make use of IQ data from the fronthaul interface in UL slots to capture the whole picture of the current channel state. In the subsequent paragraphs, we will outline our experimental approach and present our findings. 

\begin{figure}[!t]
    \centering

    \begin{subfigure}[b]{1.0\linewidth}
        \centering
        \includegraphics[width=\linewidth]{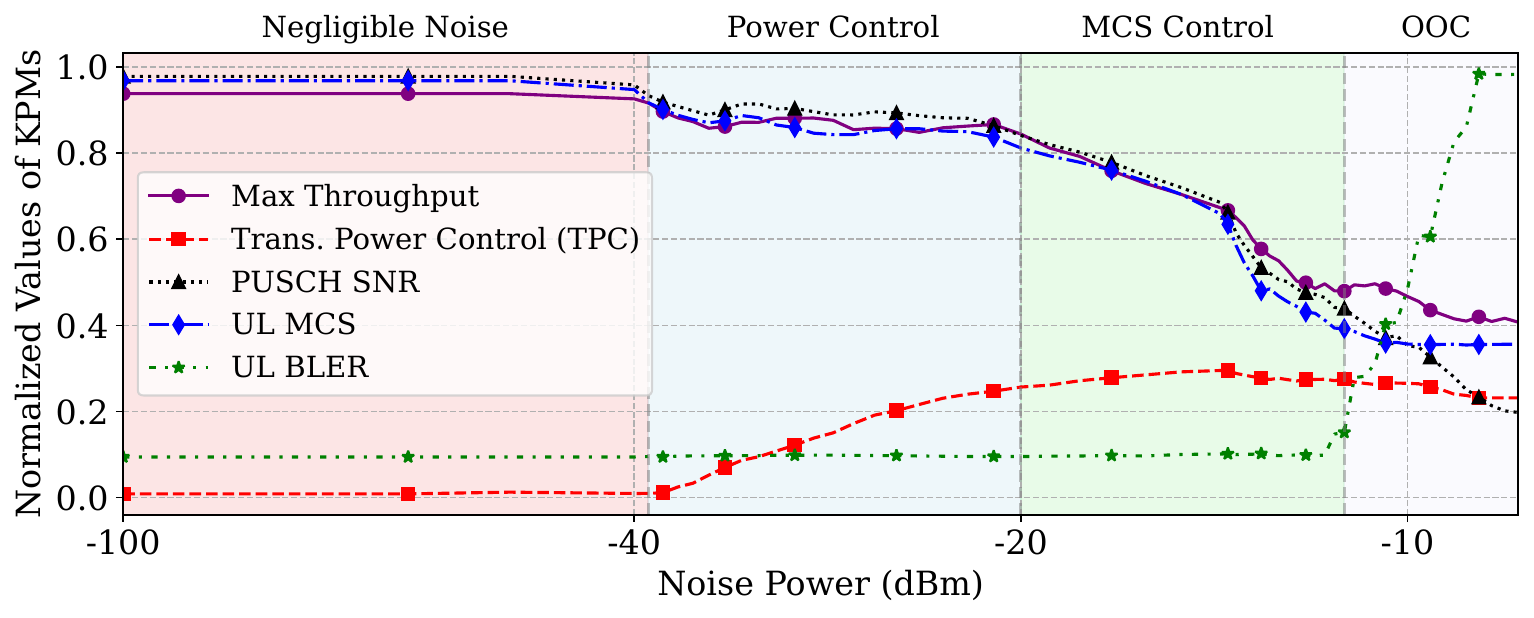}
        \caption{High-load scenario (\texttt{iPerf3} bandwidth is set to 128 Mbps)}
        \label{fig:gNB_highload}
    \end{subfigure}
   \begin{subfigure}[b]{1\linewidth}
       \centering
        \includegraphics[width=\linewidth]{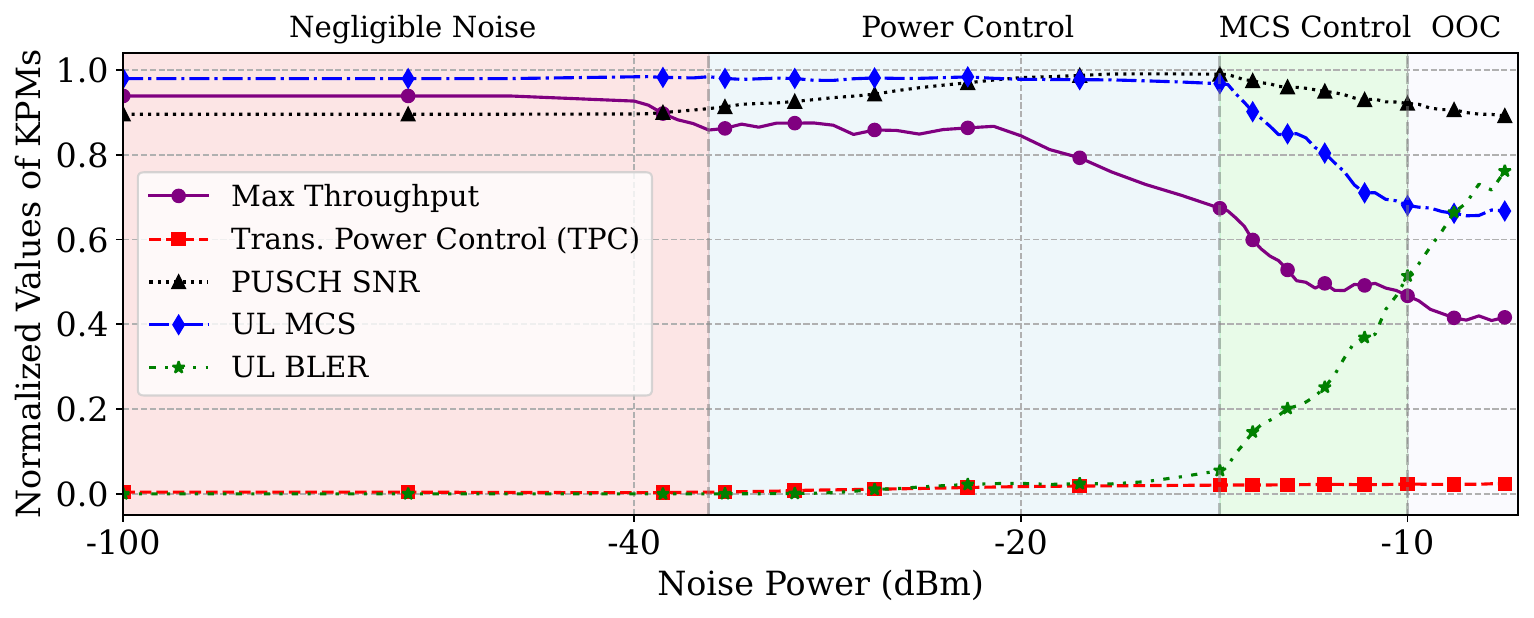}
        \caption{Low-load scenario (\texttt{iPerf3} bandwidth is set to 1 Mbps)}
        \label{fig:gNB_lowload}
   \end{subfigure}
    \begin{subfigure}[t]{0.22\linewidth}
        \centering
        \includegraphics[width=\linewidth]{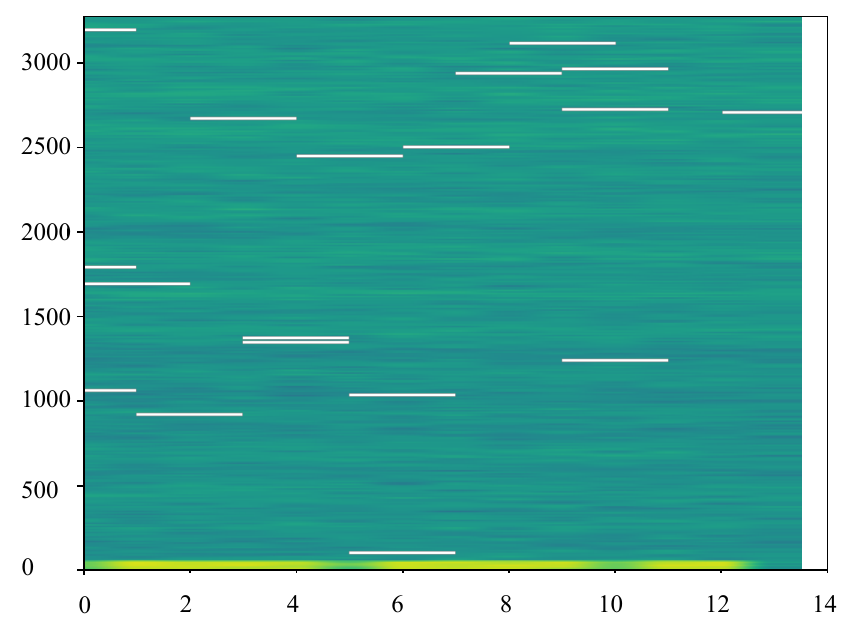}
        \caption{-60.0 dBm}
        \label{fig:Spectrogram_60}
    \end{subfigure}
    \hfill
    \begin{subfigure}[t]{0.22\linewidth}
        \centering
        \includegraphics[width=\linewidth]{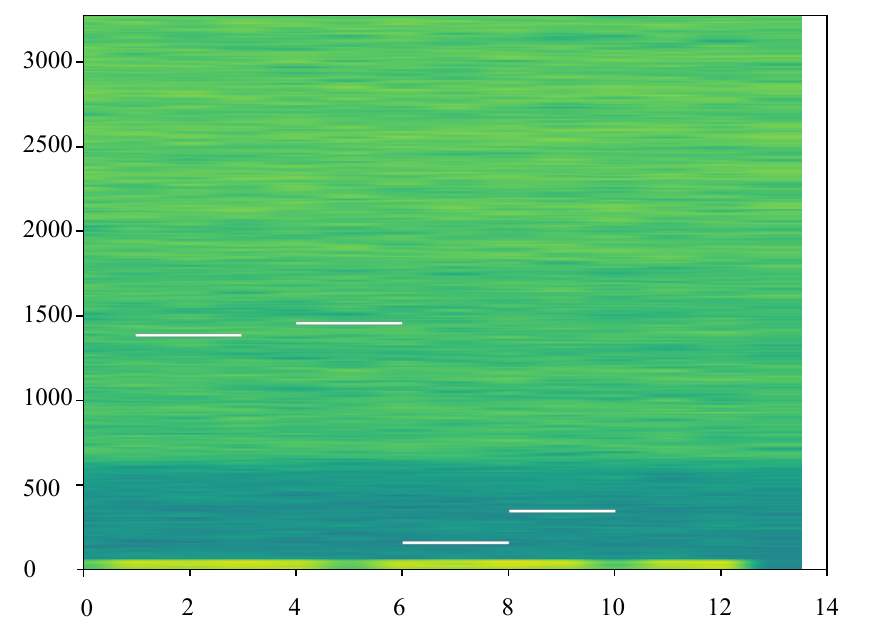}
        \caption{-30.0 dBm}
        \label{fig:Spectrogram_30}
    \end{subfigure}
    \hfill
    \begin{subfigure}[t]{0.22\linewidth}
        \centering
        \includegraphics[width=\linewidth]{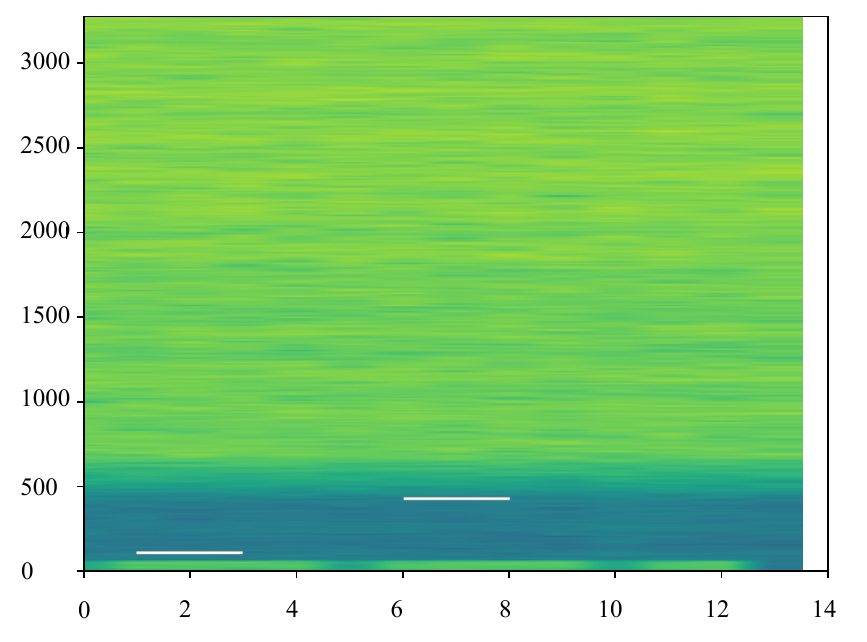}
        \caption{-15.0 dBm}
        \label{fig:Spectrogram_15}
    \end{subfigure}
    \hfill
    \begin{subfigure}[t]{0.22\linewidth}
        \centering
        \includegraphics[width=\linewidth]{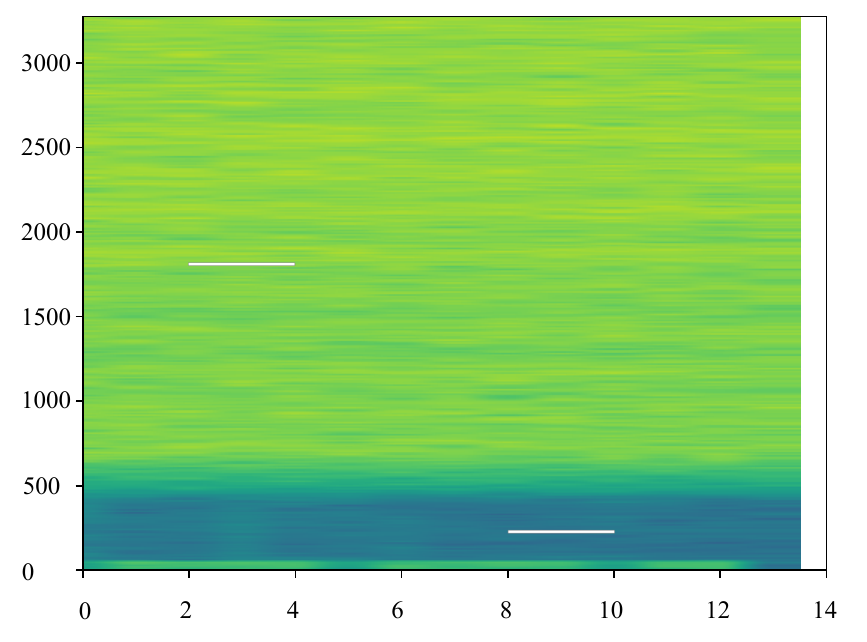}
        \caption{-9.0 dBm}
        \label{fig:Spectrogram_9}
    \end{subfigure}

    \caption{Correlation between our proposed additional KPMs and UL maximum achievable throughput under different interference levels (a,b).
    Example spectrograms of UL slots at different noise levels (c,d,e,f).}
    \label{fig:Correlation_KPM_throughput}
    \vspace{-5mm}
\end{figure}

In our experiment, we use the \texttt{iPerf3} tool to measure the maximum achievable throughput between UE and Edge server under incremental noise levels, shown as ``Max Throughput'' lines in Fig.~\ref{fig:gNB_highload} and~\ref{fig:gNB_lowload}. We then repeat the process with a change in target throughput of \texttt{iPerf3} to have two different scenarios: (1) high UL load (\texttt{iPerf3} bandwidth is set to 128 Mbps)  and (2) low UL load (\texttt{iPerf3} bandwidth is set to 1 Mbps). KPMs are also recorded for both scenarios.


Fig.~\ref{fig:gNB_highload} shows the correlation between uplink KPMs (UE's TPC, PUSCH SNR, UL MCS, and UL BLER) and UL maximum achievable throughput under varying interference levels.
When the noise power is under \textit{Negligible Noise} zone (i.e., red shaded area), the maximum UL throughput is at its peak.  
When the interference power is under UE's \textit{Power Control} zone, the gNB instructs the UE to increase transmission power (via TPC) to maintain SINR at the target level. 
However, if the interference power is increased further into \textit{MCS Control} zone (i.e., green shaded area), either the gNB RX antenna sensitivity is saturated or UE's power headroom reaches the minimum, the gNB responds by lowering the UL MCS to maintain UL BLER stable at the target value.
When the interference power further increases to the \textit{Out of Control (OOC)} zone (i.e., violet shaded area), the UL BLER rises sharply, as no further adjustments can be made to maintain signal quality or UL throughput.

However, for low-load scenario in Fig.~\ref{fig:gNB_lowload}, there is low correlation between these metrics and UL throughput. Especially under \textit{Power Control} zone, the maximum achievable throughput drops significantly while the values of KPMs remain unchanged. 
This is because KPMs are obtained based on only a small portion of PRBs allocated for transmission (the first few PRBs in  Fig.~\ref{fig:Spectrogram_60}), which are not affected by interference. 
To address this, we use the spectrogram from fronthaul IQ data (see Figs.~\ref{fig:Spectrogram_9}, \ref{fig:Spectrogram_15}, \ref{fig:Spectrogram_30}, \ref{fig:Spectrogram_60}) to capture the channel conditions.

In case that BLER reaches 100\% (see the tail of Fig.~\ref{fig:gNB_highload} under the \textit{OOC} zone), we propose to use HARQ-RV (Redundant Version) counters of a transport block (TB) to capture throughput. The BLER is calculated as the ratio between two increments: the increment of HARQ-1 over the increment of HARQ-0. Therefore, even if BLER reaches 100\%, gNB and UE can still communicate using the remaining RVs (2,3). As a result, tracking the changes of HARQ counters helps to estimate the maximum achievable throughput.



\subsection{ML Model Architecture for Throughput Estimator}
\label{subsec:Proposed_ML_throughput_Estimator}

\begin{figure}
    \centering
    \includegraphics[width=0.8\linewidth]{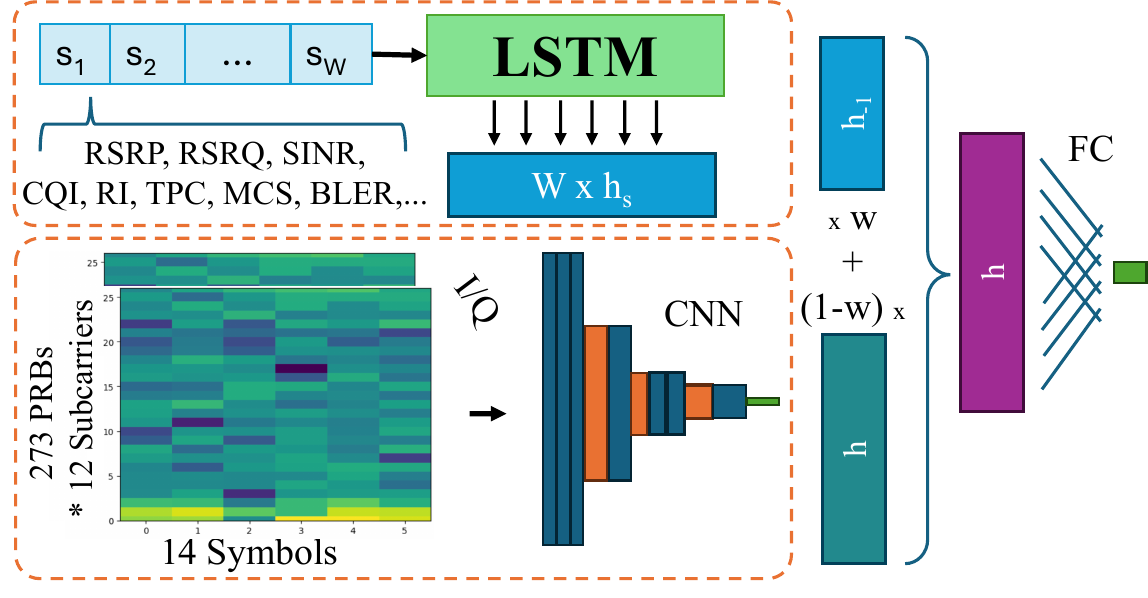}
    \caption{AI-based model architecture for throughput estimator.}
    \label{fig:throughput_estimator_model}
\end{figure}

Our input features include two categories: (i) numerical data (e.g., RSRP, RSRQ, SINR, CQI, RI, PHR, MCS, BLER, etc.) with temporal correlation, and (ii) image-based from IQ-data. 
Fig.~\ref{fig:throughput_estimator_model} shows our proposed ML model architecture for the throughput estimator, which includes two branches.
In branch 1, an LSTM model (with hidden size=124, window=30) is used to capture the temporal relation of numerical KPMs. 
In branch 2, a CNN (shown in Table~\ref{tab:cnn_architecture}) is used to extract spatial features (e.g., patterns in 2D data), 
producing a representational vector that captures throughput and wireless conditions.
Finally, we compute the weighted sum of the outputs from spatial (CNN) and temporal (LSTM) features, which are then fed into a final fully connected (FC) layer to perform a regression task that estimates throughput. The weight is the ratio of gNB-allocated resources to the UE and total available resources.

\begin{table}[!t]
\centering
\caption{Architecture of CNN model for IQ-based input.}
\label{tab:cnn_architecture}
\begin{tabular}{l l}
\toprule
 \textbf{Layers} & \textbf{Output Shape} \\
\midrule
 Input & $(N, 2, 273 \cdot 12 , 14)$ \\
 Conv2d | Relu & $(N, 16, 3276, 14)$ \\
 MaxPool2d & $(N, 16, 1638, 7)$ \\
Conv2d | Relu & $(N, 32, 1638, 7)$ \\
 MaxPool2d & $(N, 32, 819, 3)$ \\
Flatten & $(N, 32 \cdot 819 \cdot 3)$ \\
 Linear | Relu | Dropout & $(N, \text{hidden\_size})$ \\
\bottomrule
\end{tabular}
\vspace{-5mm}
\end{table}

\section{Adaptive ML Partitioning Algorithm}

Prior work typically optimizes a single metric, such as latency, as the primary objective, overlooking the interplay of multiple considerations. In contrast, we propose a comprehensive approach that jointly considers three critical metrics—latency, privacy, and energy.
In Adaptive ML splitting enabled by throughput estimator over dynamic 5G networks, optimising the following three objective functions is critical for efficient and privacy-preserving inference offloading:
\begin{itemize}
    \item \textbf{E2E delay ($D_{E2E}$)}: The total time required for a given inference task, including (i) local computing at the UE $d_{\text{UE}}$, (ii) transmission latency of intermediate features $d_{\text{TRX}}$, and (iii) remote inference time at the edge/cloud $d_{\text{ser}}$; $D_{E2E}(l, TP)=d_{\text{UE}}(l) +d_{\text{TRX}} (l, TP) +d{\text{ser}} (l)$, where $TP$ is the bandwidth between the UE and the Edge. 
    \item \textbf{Privacy Leakage ($P$)}: The sensitivity of the data, quantified using distance correlation $\rho$ \cite{nopeek} between the original input image and the transmitted intermediate layer. \textit{Lower values of $\rho$ indicate better privacy protection}.
    \item \textbf{UE's energy consumption ($E_{\text{UE}}$)}: The energy consumed by the UE to process the Head model before transmitting the intermediate features. For a CPU, this is computed using the power drawn during inference, with the thermal design power (TDP) per thread serving as the average power consumption, $E_{\text{UE}} = TDP / \text{threads} \cdot t$. 
\end{itemize}

Our objective is to \textit{jointly} optimize these three metrics by \textit{determining the optimal splitting point $l^*$} of the ML model between the UE and edge/cloud, 
that minimizes the weighted sum of E2E latency, privacy risk, and energy consumption:
\begin{equation}
    l^* = \arg \min_{l \in \{1, \dots, L\}}  F(l,TP)
\end{equation}
where the objective function is :
\begin{equation}
     F(l,TP) = w_1 D_{\text{E2E}}(l, TP) + w_2 P(l) + w_3 E_{\text{UE}}(l)
     \label{eq:ObjectiveFunction}
\end{equation}
 
and weighting parameters \( w_1 \), \( w_2 \), and \( w_3 \) are designed to normalize these metrics, ensuring balanced contributions to \( F(l, TP) \).
The problem is subject to the following constraints:
\begin{align}
    D_{\text{E2E}} &\leq \tau_{\max}
    \quad \text{(Latency Constraint)} \nonumber \\
    P = \rho &\leq \rho_{\max}
    \quad \text{(Privacy Constraint)} \nonumber \\
    E_{\text{UE}} &\leq E_{\max}
    \quad \text{(Energy Constraint)} \nonumber
\end{align}


In a dynamic 5G environment, fluctuating throughput \( TP \) reported by the RAN poses challenges for maintaining acceptable latency, privacy protection, and energy efficiency while optimizing inference performance. The frequent variations in \( TP \) necessitate rapid selection of the optimal splitting point \( l^* \). 
We construct a lookup table that precomputes and stores \( l^* \) for each (rounded in Mbps) \( TP \in \{1, \dots, TP_{max}\}\), where $TP_{max}$ is the maximum theoretical UE's throughput. This allows the \textit{AF} to retrieve \( l^* \) in constant time.

We propose the Pre-filtered Split Optimization (PSO) algorithm, detailed in Algorithm~\ref{alg:PFBSO}.
For each UE, given the maximum achievable throughput reported by the RAN's Throughput Estimator (proposed in Sec.~\ref{subsec:Proposed_ML_throughput_Estimator}), the \textit{AF} can efficiently select the optimal splitting point using the corresponding lookup table obtained from Algorithm~\ref{alg:PFBSO}.

\SetAlFnt{\small}
\begin{algorithm}[!t]
\caption{Pre-Filtered Split Optimization (PSO)}
\label{alg:PFBSO}
\KwIn{
Set of UEs $\mathcal{U}$, candidate splitting points $l \in \{1, \dots, L_u\}$, each with: privacy $P_u(l)$, energy $E_{\text{UE},u}(l)$, processing delays $d_{\text{UE,u}}(l), d_{\text{ser}}(l)$, intermediate data size $\text{data\_size}_u(l)$; \\
UE throughput range $TP_u \in \{1, \dots, TP_{\max,u}\}$; \\
Constraints: $\tau_{\max,u}$, $\rho_{\max,u}$, $E_{\max,u}$
}
\KwOut{A set of $\{\text{LookupTable}_u\}_{u \in \mathcal{U}}$  mapping throughput $TP_u$ to optimal splitting point $l^*$.}

\For{ each UE $u \in \mathcal{U}$ }{
Initialize feasible set $\mathcal{L} \gets \emptyset$ 

\For{$l \in \{1, \dots, L_u\}$}{
  \If{$P_u(l) \leq \rho_{\max,u}$ \textbf{and} $E_{\text{UE},u}(l) \leq E_{\max,u}$}{
    Compute minimal required throughput $TP_{\min,u}(l) \gets \dfrac{\text{data\_size}_u(l)}{\tau_{\max,u} - d_{\text{UE,u}}(l) - d_{\text{ser}}(l)}$\\
      $\mathcal{L} \gets \mathcal{L} \cup \{(l, TP_{\min,u}(l))\}$
  }
}

Initialize $\text{LookupTable}_u$ $\gets \emptyset$

\For{$TP \in \{1, \dots, TP_{\max,u}\}$}{
  Identify feasible splits at this $TP$: \\
$\hat{\mathcal{L}} \gets \{l \mid (l, TP_{\min,u}(l)) \in \mathcal{L},\ TP_{\min,u}(l) \leq TP\}$\\

  \If{$\hat{\mathcal{L}} \neq \emptyset$}{
    Select $l^* \gets \arg\min_{l \in \hat{\mathcal{L}}} F(l, TP)$ \\
    Add $(TP, l^*)$ to $\text{LookupTable}_u$
  }
}
Add $\text{LookupTable}_u$ to $\{\text{LookupTable}_u\}$
}
\Return $\{\text{LookupTable}_u\}_{u \in \mathcal{U}}$
\end{algorithm}

\section{Experimental Setup and Results}
\label{sec:ExperimentalResults}

\subsection{Experimental Setup}
\begin{figure}[!t]
    \centering
    \begin{subfigure}[b]{0.32\linewidth}
        \centering
        \includegraphics[width=\linewidth]{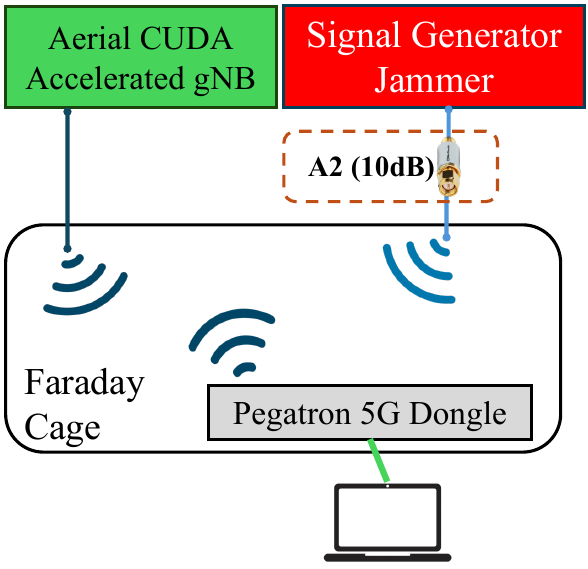}
        \caption{S1: Jamming}
        \label{fig:sce1}
    \end{subfigure}
    \hfill
    \begin{subfigure}[b]{0.31\linewidth}
        \centering
        \includegraphics[width=\linewidth]{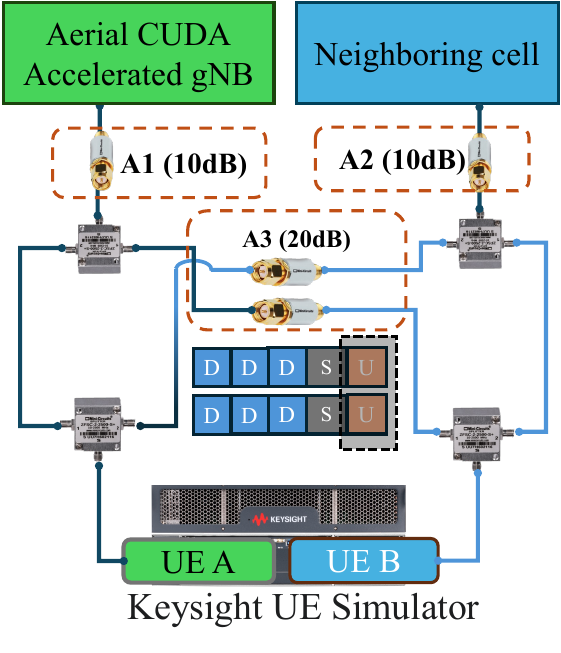}
        \caption{S2: UE-to-BS}
        \label{fig:sce2}
    \end{subfigure}
    \hfill
    \begin{subfigure}[b]{0.31\linewidth}
        \centering
        \includegraphics[width=\linewidth]{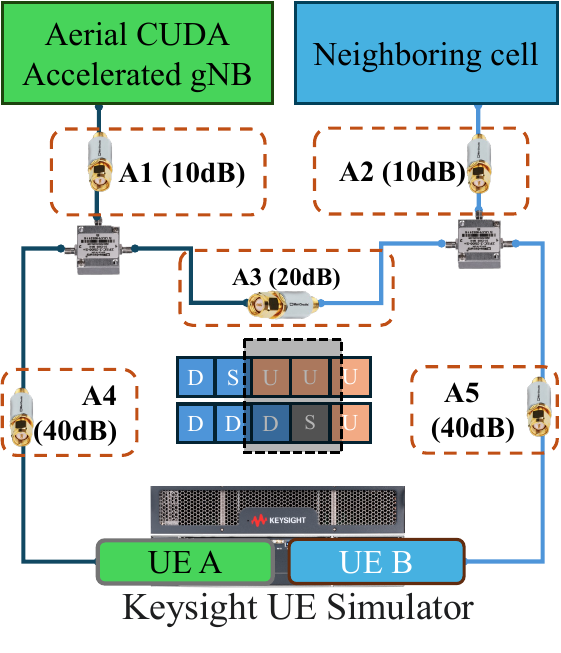}
        \caption{S3: BS-to-BS}
        \label{fig:sce3}
    \end{subfigure}

    \caption{Setup of different interference scenarios.}
    \label{fig:interference_scenarios}
\end{figure}

For AI-based image analysis, we used VGG16 model~\cite{simonyan2014vgg}, known for its simplicity and strong classification performance, and divided it into two components for optimized deployment over 5G network:

\begin{itemize}
    \item \textbf{Head model}: Deployed on the 5G UE. A virtual machine with 4GB RAM and 2 CPU cores simulates the limited resources of a mobile device. This VM is connected—via pass-through mode—to either a Pegatron 5G dongle or the Keysight UE-SIM to access the 5G wireless channel.

    \item \textbf{Tail model}: Hosted on a high-performance GPU server with 512GB of RAM, and 2x NVIDIA A40 GPUs.
\end{itemize}
VGG16 offers 43 potential splitting points, corresponding to outputs after each layer or sub-layer (e.g., convolutions, ReLU, or pooling). These points enable analysis of intermediate outputs, including their size and distance correlation with the input. For each splitting point $l$, we processed 200 images from the ImageNet dataset to evaluate characteristics of $P(l)$ and $E_\text{UE}(l)$.
To evaluate the performance of the proposed Adaptive Splitting mechanisms (hosted in AF), we conducted experiments under normal scenario and three interference scenarios: 

    
1) \textbf{S1: Jamming Interference:}  
    As in Fig.~\ref{fig:sce1}, a Signal Generator acts as a jammer. The uplink (UL) signal transmitted from the UE to the BS was affected by abnormal external interference, leading to significant UL performance degradation.

2) \textbf{S2: UE-to-BS Interference (Co-Channel Interference):}  
    As in Fig.~\ref{fig:sce2}, an UL from a UE to its serving BS was influenced by another UE connected to a nearby BS. The unwanted signal introduced interference in the uplink of the primary UE, leading to \textit{co-channel interference (CCI)}.

3) \textbf{S3: BS-to-BS Interference (TDD Pattern Mismatch):} 
    As in Figure~\ref{fig:sce3}, the right-hand interfering cell uses a heavy DL profile, while the left-hand victim cell is optimized for UL-dominant use cases. 
    The mismatch in TDD patterns resulted in overlapping downlink (D) and uplink (U) time slots, leading to UL performance degradation in the victim cell.

\subsection{Results and Analysis for Throughput Estimator}
Table~\ref{tab:model_performance} presents the $R^2$ score and $RMSE$ of three ML models in low-load scenario under interfered environment with different kinds of input features.
As shown in the table, we significantly enhanced the performance of XGBoost by adding eight features.
In our proposed Model C, we combine representations of KPMs and IQ data, and achieve the best $R^2$ score of 0.9636, and the lowest $RMSE$ value on the test set.

\begin{table}[!t]
\centering
\caption{Model Performance Comparison}
\label{tab:model_performance}
\begin{tabular}{llcc}
\toprule
\textbf{Model} & \textbf{Input features} & $R^2$ & $RMSE$ \\
\midrule
A: XGBoost~\cite{Minovski_Thrput_TMC_2023} & 7 KPMs \cite{Minovski_Thrput_TMC_2023} & 0.3160 & 10.7748 \\
B: XGBoost & 15 KPMs (+8 proposed KPMs) & 0.7845 & 6.0478 \\
C: Proposal & Timeseries (15 KPMs) + IQ  & \textbf{0.9636} & \textbf{2.4839} \\
\bottomrule
\end{tabular}
\vspace{-5mm}
\end{table}


\subsection{Results for Adaptive AI Splitting Algorithm}


We first analyze single metric—E2E delay, Privacy, or UE's energy consumption of the objective function, and how they contribute to the joint objective function. This approach provides deeper insights into model behaviour when it is split.

\begin{figure*}[htbp]
    \centering
    \begin{subfigure}[b]{0.24\linewidth}
        \centering
        \includegraphics[width=\linewidth]{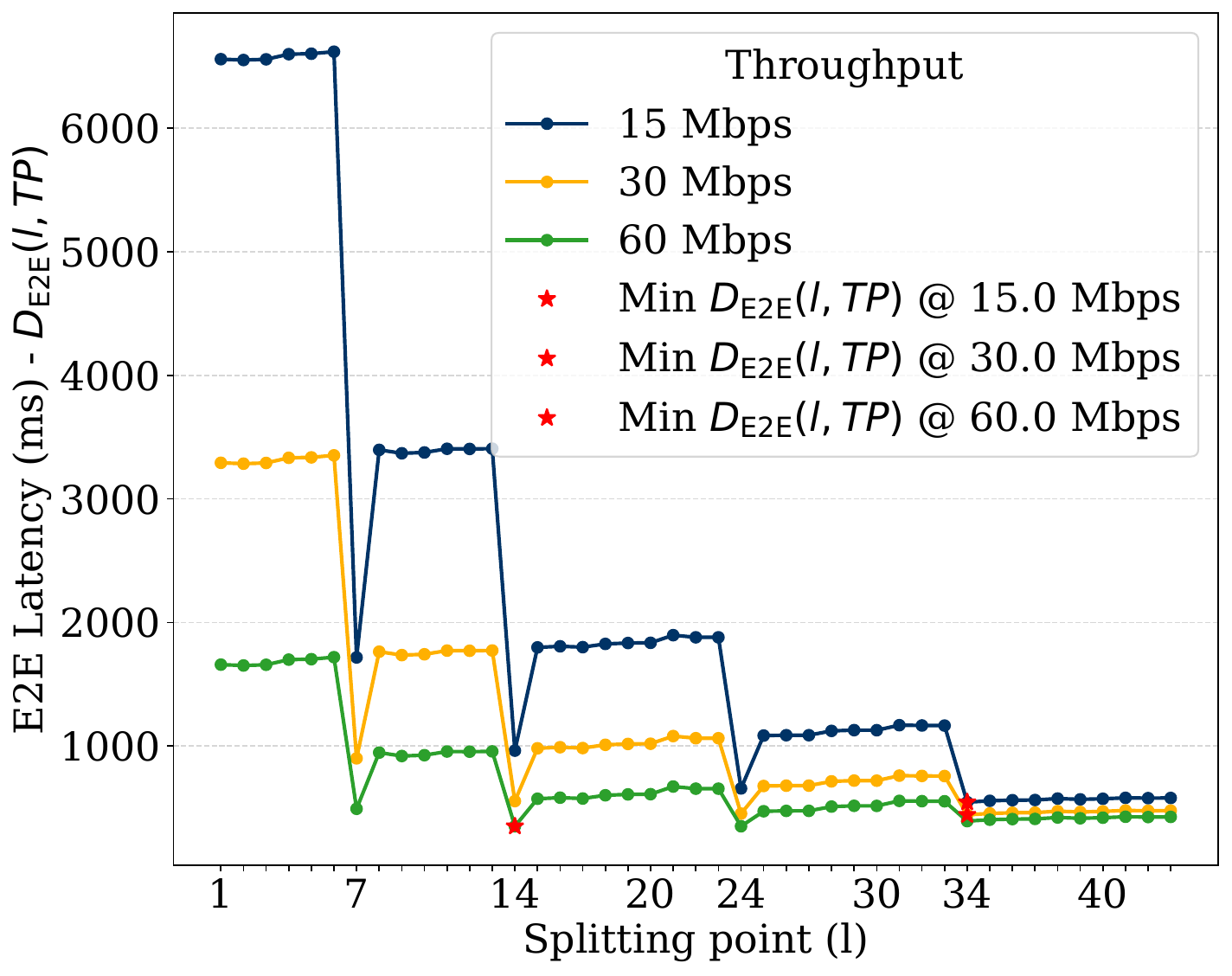}
        \caption{E2E delay $D_\text{E2E}(l, TP)$}
        \label{fig:F_vs_splittingPoint_latency_focus}
    \end{subfigure}
    \hfill
    \begin{subfigure}[b]{0.24\linewidth}
        \centering
        \includegraphics[width=\linewidth]{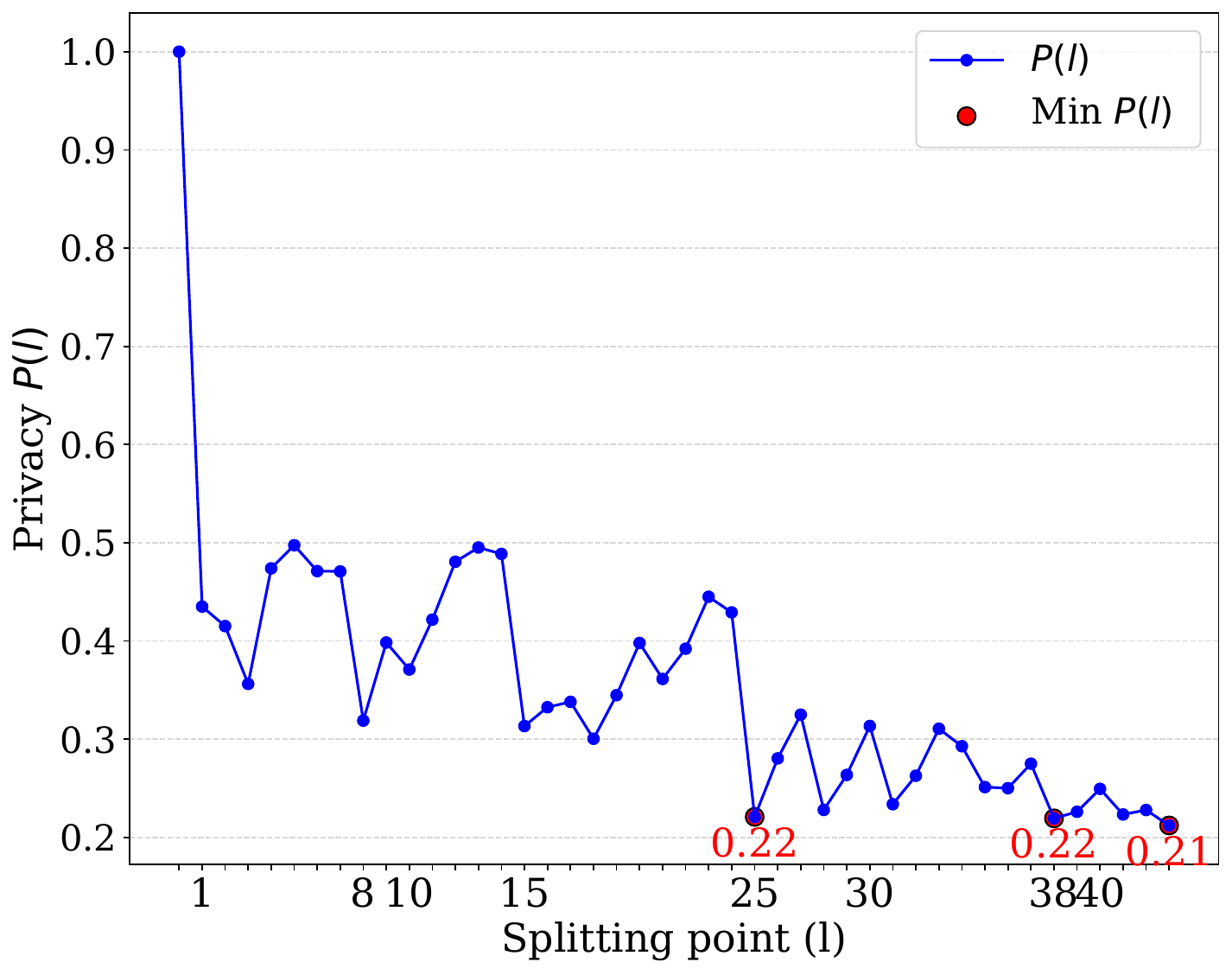}
        \caption{Privacy Leakage $P(l)$}
        \label{fig:F_vs_splittingPoint_privacy_focus}
    \end{subfigure}
    \hfill
    \begin{subfigure}[b]{0.24\linewidth}
        \centering
        \includegraphics[width=\linewidth]{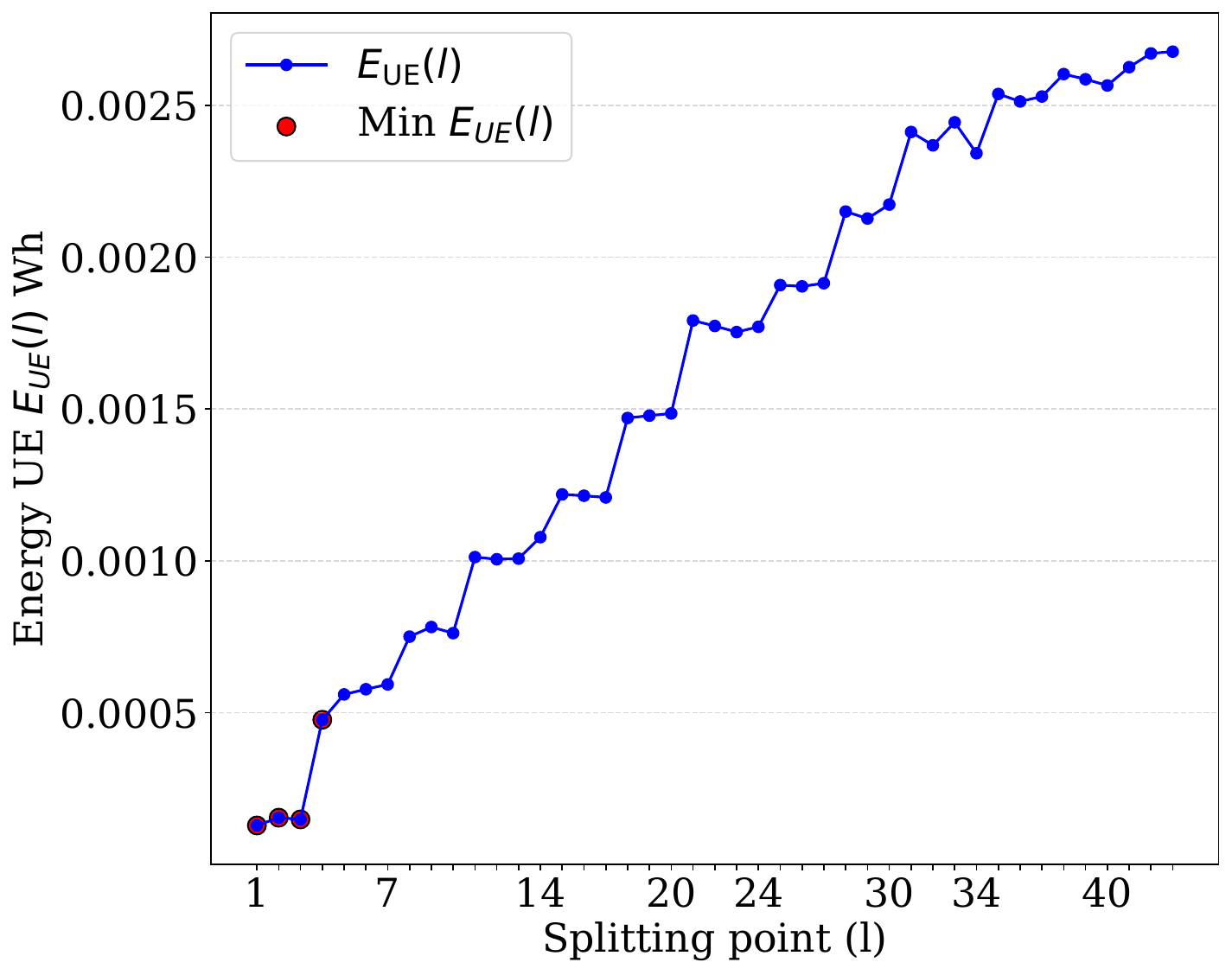}
        \caption{UE's Energy $E_\text{UE}(l)$}
        \label{fig:F_vs_splittingPoint_energy_focus}
    \end{subfigure}
    \hfill
    \begin{subfigure}[b]{0.24\linewidth}
        \centering
        \includegraphics[width=\linewidth]{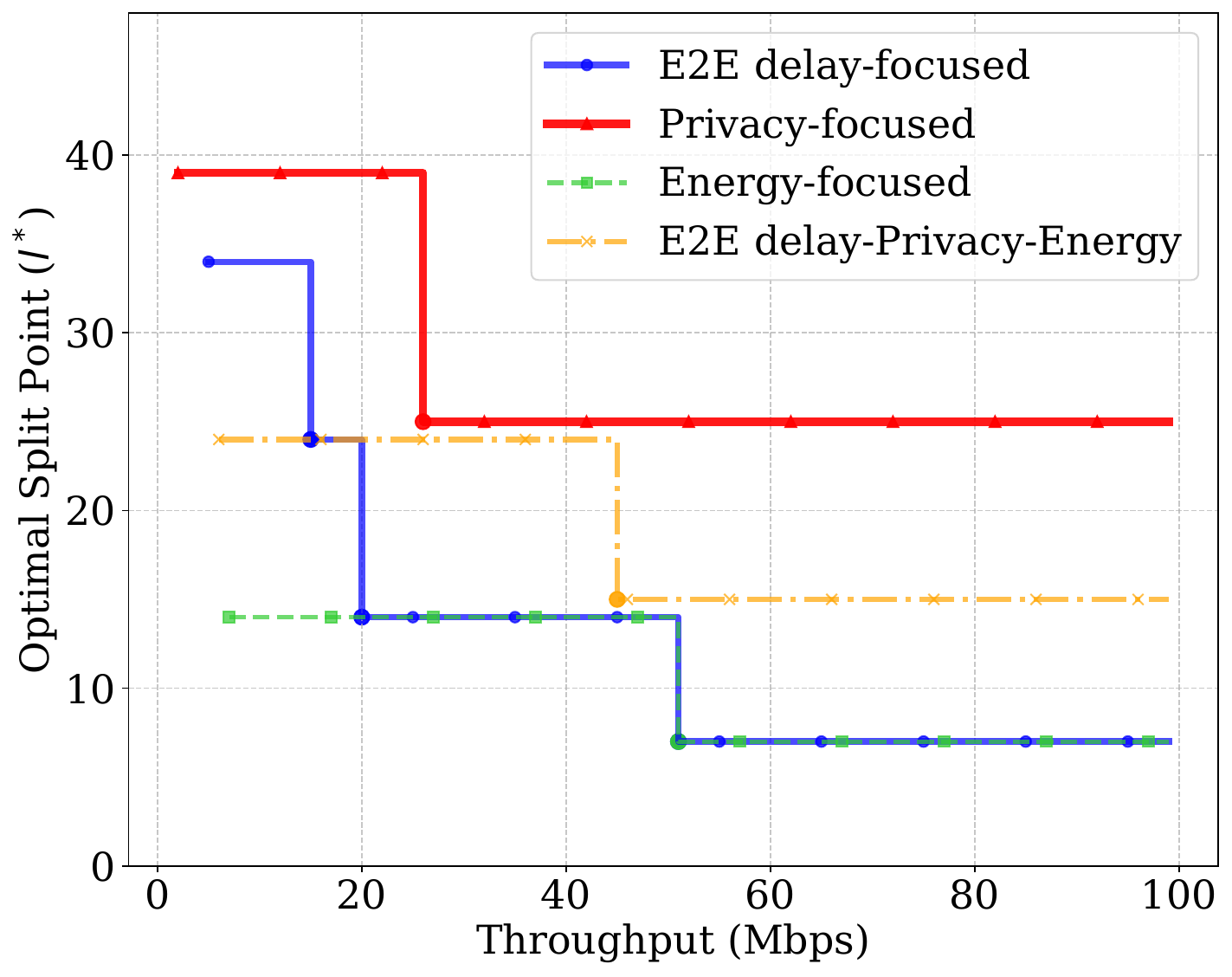} %
        \caption{Jointly optimize three metrics}
        \label{fig:F_vs_l_combination}
    \end{subfigure}

    \caption{Objective function $F(\ell, TP)$ vs. splitting point $\ell$ under various optimization focuses, i.e., different weighting parameters.}
    \vspace{-5mm}\label{fig:combined_F_vs_splittingPoint}
\end{figure*}

Under the \textit{E2E delay-only} objective ($w_1=1, w_2=w_3=0$), we simulate three throughput conditions: (i) 60 Mbps---excellent signal quality; (ii) 30 Mbps---stable network condition; and (iii) 15 Mbps---degraded signal quality or high interference. As shown in Fig.~\ref{fig:F_vs_splittingPoint_latency_focus}, 
occasional drops in latency can be observed at certain splitting points. These are primarily attributed to the data output from \textit{MaxPooling} layers, which reduce the feature map size and consequently lower the transmission cost. The red stars indicate the splitting points where the E2E latency is minimized for each throughput condition.
        
Under the \textit{Privacy-only} objective, 
Fig.~\ref{fig:F_vs_splittingPoint_privacy_focus} shows a clear downward trend of privacy leakage; the privacy score is highest near the input layers and gradually decreases as the model is split deeper. A sharp decline occurs around splitting point \textit{25}, with further reductions and stabilization observed in the later stages. The minimum privacy leakage highlighted by red-circled points, appear at splitting points \textit{25}, \textit{38}, and \textit{43}, where values drop to approximately 0.21–0.22. This indicates that deeper splitting points are more effective at preserving privacy, likely due to the increasingly abstract nature of the features, which reduces the risk of leaking sensitive input information.

Under the \textit{Energy-only} objective
,Fig.~\ref{fig:F_vs_splittingPoint_energy_focus}, the Energy of UE increases consistently as the splitting point moves deeper, indicating higher energy consumption. The lowest values, highlighted by red circles, occur at splitting points $1$, $2$, and $3$, suggesting that early splits are the most energy-efficient.



    To jointly optimize the objective function for different strategies, we set the target constraints $\tau_{\max}$, $\rho_{\max}$, and $E_{\max}$ according to the specific metric we aim to prioritize. Figure~\ref{fig:F_vs_l_combination} illustrates how the optimal splitting point varies depending on the chosen optimization focus.  
   In the \textit{E2E delay-focused} case, the optimal splitting point shifts progressively from around 7 to 14, then 24, and finally to 34 as the throughput decreases, which aligns with the \textit{red star} points in Fig.~\ref{fig:F_vs_splittingPoint_latency_focus}. 
   The \textit{Privacy-focused} strategy tends to keep the splitting point deeper in the ML model. 
   In contrast, \textit{energy-focused} strategy (i.e., green dotted line) pushes the splitting point earlier to reduce on-device computation and conserve energy.
   Finally, the \textit{E2E delay-Privacy-Energy} combination strategy finds a middle ground between all three metrics, resulting in a dynamic but balanced shift of the optimal splitting point.
   Under very low throughput conditions, some strategies fail to find the optimal point, as shown missing early part of some lines in Fig.~\ref{fig:F_vs_l_combination}.
    
\begin{figure}
    \centering
    \includegraphics[width=0.8\linewidth]{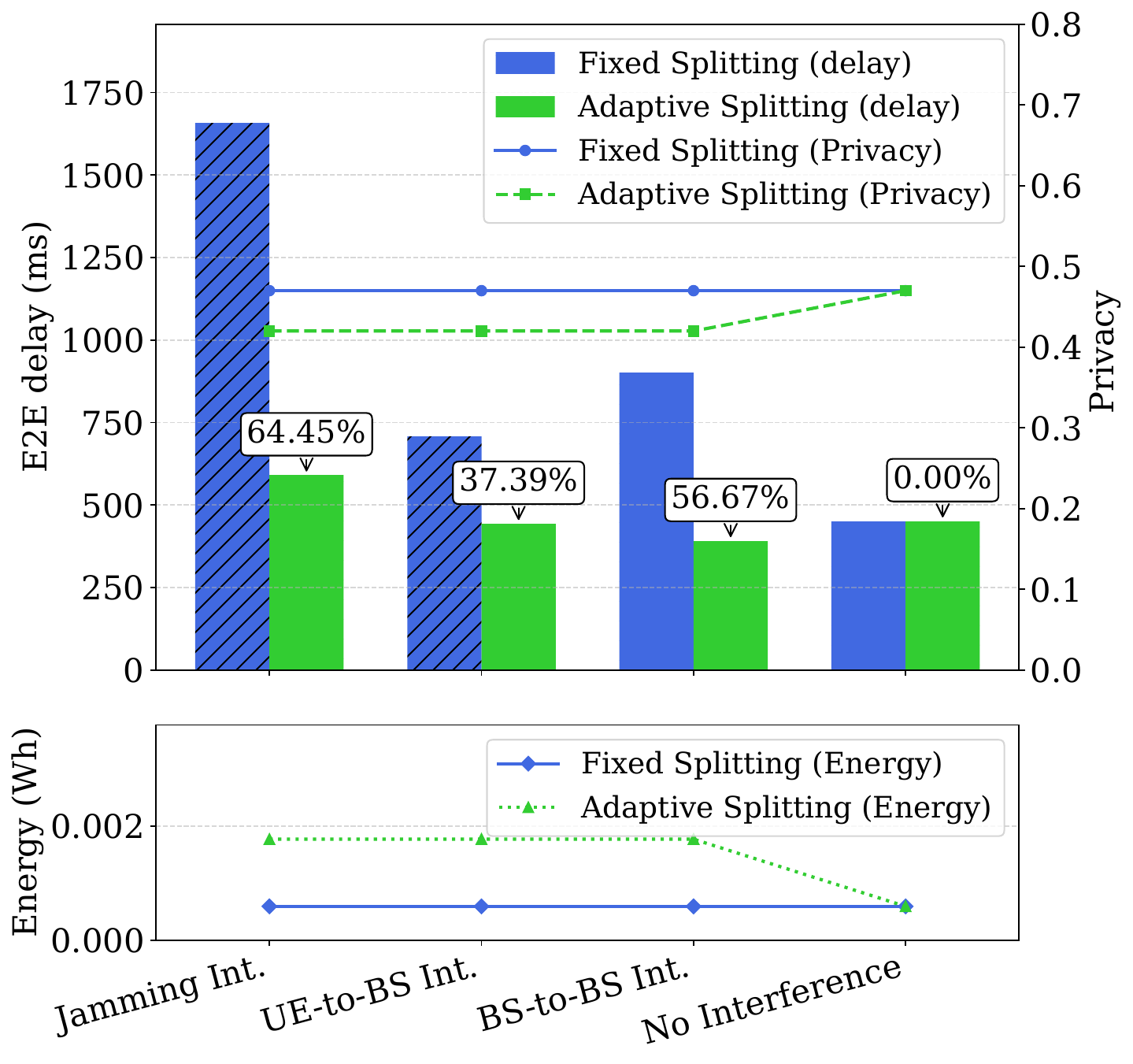}
    \caption{Comparison between Fixed vs. Adaptive Splitting under different scenarios, in terms of E2E delay, Privacy, UE Energy.}
    \vspace{-5mm}
    \label{fig:FixedVsAdaptive}
\end{figure}

    Fig.~\ref{fig:FixedVsAdaptive} shows a comparison between Fixed Splitting and Adaptive Splitting under different wireless conditions, in terms of three metrics.  
    We can see under \textit{No Interference} scenario, the two methods obtained the same performance on three metrics.
    However, under three interference scenarios, Adaptive Splitting significantly reduces E2E delay  with minimal impact on privacy, but the trade-off is 
    an increase of UE energy. 
    Under \textit{Jamming} scenario, E2E delay decreases from 1,657\,ms to 589\,ms, achieving \textbf{64.45\%} reduction. For \textit{UE-to-BS Int.} and \textit{BS-to-BS Int.} scenarios, E2E delay is improved by 37.39\% and 56.67\%, respectively.
    In summary, the proposed Adaptive Splitting is highly effective in reducing E2E delay under interference conditions while incurring an increase in UE Energy due to more computation on UE. 
%


\section{Conclusion}
\label{sec:conclusion}
In this paper, we have presented a comprehensive ML-partitioning algorithm jointly optimizing three objectives (privacy, E2E delay, energy consumed by UE). We also investigated different input features for throughput estimators under challenging wireless environment, and include IQ-based spectrogram image to the proposed ML-based throughput estimator. 
Our evaluation of the proposed solution on a testbed running NVIDIA Aerial 5G stack demonstrates promising results. 

\textbf{Acknowledgment}: 
This research is supported in part by the National Research Foundation (NRF), Singapore and Infocomm Media Development Authority (IMDA) under its Future Communications Research \& Development Programme, and in part by the SNS JU project 6G-GOALS under the EU's Horizon program (Grant Agreement No. 101139232). Any opinions, findings, conclusions, or recommendations expressed in this material are those of the authors and do not reflect the views of NRF and IMDA Singapore, or the European Union.

\bibliographystyle{IEEEtran}
\bibliography{references}

\end{document}